# Temporal variability and statistics of the Strehl ratio in adaptive-optics images[1]

Szymon Gladysz

*Department of Experimental Physics, National University of Ireland, Galway, University Road, Galway, Ireland*

szymon.gladysz@nuigalway.ie

Julian C. Christou

*Division of Astronomical Sciences, National Science Foundation, Arlington, VA, USA* [2]

jchristo@nsf.gov

L. William Bradford

lawrence.w.bradford@boeing.com

Lewis C. Roberts, Jr.[3]

lewis.c.roberts@jpl.nasa.gov

*The Boeing Company, Kihei, HI, USA*

ABSTRACT

We have investigated the temporal variability and statistics of the "instantaneous" Strehl ratio. The observations were carried out with the 3.63-m AEOS telescope equipped with a high-order adaptive optics system. In this paper Strehl ratio is defined as the peak intensity of a single short exposure. We have also studied the behaviour of the phase variance computed on the reconstructed wavefronts. We tested the Maréchal approximation and used it to explain the observed negative skewness of the Strehl ratio distribution. The estimate of the phase variance is shown to fit a three-parameter Gamma

---

distribution model. We show that simple scaling of the reconstructed wavefronts has a large impact on the shape of the Strehl ratio distribution.

1. INTRODUCTION

In this paper, we analyze the properties of the normalized on-axis intensity, i.e. the Strehl ratio (SR), of the adaptive optics (AO) short-exposure images. This research was prompted by the premise of performing frame selection or "lucky imaging" in order to increase the quality of astronomical images *after* AO. We recently showed (Gladysz et al. 2007) that the knowledge of the probability density function (PDF) of SR completely determines the outcome of the method, i.e. SR and the signal-to-noise ratio of the final shift-and-add image. Therefore - given an analytical form of the PDF - we can propose guidelines for applying frame selection.

Histograms of the instantaneous AO-corrected SR were first published by Gladysz et al. (2006) based on the observations with the 3-m Shane telescope at the Lick Observatory. The plots all exhibit negative skewness, i.e. low-end tail. This means fewer high-quality outliers than in the case of positively-skewed PDF which was encountered during "lucky imaging" speckle campaign by Baldwin et al. (2001). Switching AO on transforms the morphology of the SR distribution.

This effect was explained by decomposing AO phase variance into three error terms and using the Maréchal approximation to relate SR and phase variance. This approach was valid for the AO correction regime corresponding to the Lick data (mean SR of 0.4 - 0.5). In this paper we test the accuracy of the Maréchal approximation and use the distribution of phase variance to explain the morphology of the SR distribution. The observations were carried out with the Advanced Electro-Optical System (AEOS) telescope. Its 941-actuator AO system provides advanced image compensation necessary to test the intermediate-SR regime in the visible band. Since there was no short-exposure capability in the science camera the images were numerically generated from the reconstructed wavefronts sensed after the deformable mirror. As such, they were free of non-common-path errors and readout noise. It should also be mentioned that these PSFs corresponded to the estimated wavefronts, and the atmospheric aberrations which were not measured by the wavefront sensor were not present in the images.

We describe the observational setup and data analysis in § 2. The temporal variability of the two quantities we were interested in, SR and phase variance, is analyzed in § 3. The range of validity of the Maréchal approximation is examined in § 4. An analytical proof for the proposed Gamma distribution model of the phase variance estimate is given in § 5. In § 6 we give the explanation for the negative skewness of the SR histograms. In § 7 we summarize the results and present the relevant aspects of this work to high-resolution imaging.

2. OBSERVATIONS AND DATA ANALYSIS

Observations with the AEOS telescope were carried out over six nights in 2004 and 2005 using the wavefront sensor (WFS) of the 941-actuator AO system (Roberts & Neyman 2002). The WFS uses a 128x128 pixel CCD, observes between 540-700nm and can record images at up to 2 kHz. Pixel data is binned to form 32x32 subapertures, and the centroids of the images in these subapertures are used to calculate the slopes of the residual wavefront errors. This slope data is used for the next wavefront correction, and can be saved to a buffer memory. Up to 4096 such sets of slope data can be recorded.

We looked at bright stars close to zenith to take advantage of the best possible AO compensation. The observations are summarized in Table 1. Each dataset (denoted by PSF 1, 2, etc.) corresponds to one star, and typically contains 1000-2000 frames of wavefront slopes. The WFS slope data were obtained at various frame rates ranging between 200 and 2500Hz. Using software that mimics the AEOS wavefront reconstruction algorithms, estimated wavefronts could be computed for each frame of slope data.

Table 1 Point source observations with the AEOS telescope. $r_0$ (adjusted to 0.5µm and to zenith) was estimated using the differential image motion monitor (Bradley et al. 2006).

| PSF No. | Bright Star Catalogue ID | Date | $m_V$ | <SR> | Frame Rate (Hz) | $r_0$ (cm) |
|---|---|---|---|---|---|---|
| 1 | HR 5235 | 2004-04-30 | 2.68 | 0.56 | 1000 | 22 ± 5 |
| 2 | HR 8571 | 2004-04-30 | 3.75 | 0.51 | 1000 | 20 ± 5 |
| 3 | HR 5200 | 2004-05-10 | 6.04 | 0.49 | 1000 | 11 ± 8 |
| 4 | HR 7776 | 2004-05-10 | 3.08 | 0.24 | 1000 | 11 ± 8 |
| 5 | HR 6713 | 2004-07-04 | 4.67 | 0.5 | 1200 | 11 ± 3 |
| 6 | HR 7942 | 2004-07-05 | 4.22 | 0.56 | 2000 | 13 ± 3 |
| 7 | HR 6713 | 2004-08-18 | 4.67 | 0.53 | 200 | 10 ± 5 |
| 8 | HR 6713 | 2004-08-18 | 4.67 | 0.52 | 200 | 10 ± 5 |
| 9 | HR 6008 | 2004-08-18 | 5.00 | 0.53 | 200 | 10 ± 5 |
| 10 | HR 6008 | 2004-08-18 | 5.00 | 0.53 | 200 | 10 ± 5 |
| 11 | HR 6498 | 2004-08-18 | 4.35 | 0.54 | 200 | 10 ± 5 |
| 12 | HR 6498 | 2004-08-18 | 4.35 | 0.54 | 200 | 10 ± 5 |
| 13 | HR 6872 | 2004-08-18 | 4.33 | 0.47 | 200 | 10 ± 3 |
| 14 | HR 7744 | 2004-08-18 | 4.52 | 0.49 | 200 | 10 ± 3 |
| 15 | HR 7744 | 2004-08-18 | 4.52 | 0.45 | 200 | 10 ± 3 |
| 16 | HR 8775 | 2005-10-05 | 2.42 | 0.38 | 2500 | 16 ± 4 |
| 17 | HR 8775 | 2005-10-05 | 2.42 | 0.27 | 2500 | 16 ± 4 |
| 18 | HR 8775 | 2005-10-05 | 2.42 | 0.3 | 2500 | 16 ± 4 |
| 19 | HR 8775 | 2005-10-05 | 2.42 | 0.23 | 200 | 16 ± 4 |
| 20 | HR 8775 | 2005-10-05 | 2.42 | 0.06 | 200 | 16 ± 4 |
| 21 | HR 8893 | 2005-10-05 | 5.08 | 0.17 | 800 | 12 ± 4 |

At the time of this writing, there is some uncertainty in the precise calibration of the AEOS phase variances. We have scaled the data to be consistent with the range of AEOS Strehl ratios measured by other means, including our science camera. We emphasize though that *we show the SR for the telescope and AO system, rather than for any particular instrument*. Thus, if we were to look at the SR measured for the AEOS Visible Imager camera, we would have to multiply our values (at the appropriate wavelength) by about 0.7. Other science sensors such as the AEOS Spectral Imaging Sensor system (Blake et al. 2006) have different SR.

Data reduction consisted of the following steps. First, the residual global tip and tilt were removed from the wavefronts. Secondly, the average wavefront was subtracted from each wavefront in a dataset in order to make the phase spatially stationary (the importance of this step is shown in § 5). The wavefronts were then zero-padded and numerically propagated through the obscured aperture (D = 3.6m; secondary mirror's spiders were not simulated) to generate the corresponding PSFs at 0.625µm, the central wavelength of the observing passband. The amount of zero-padding was chosen as to ensure a focal-plane sampling of $\lambda/4D$.

Phase variance was estimated from the reconstructed phase maps after de-tilting. This was done in order to allow for comparison between the computed phase variance and the instantaneous SR through the Maréchal approximation. We measured SR as the peak image intensity, as opposed to the on-axis intensity. This is because we were looking for a shift-invariant image quality metric (in the case of short-exposures residual image motion can be removed by accurately registering the frames).

Computing the Strehl ratio for the AEOS data was done by comparing the frames with the numerically obtained diffraction-limited PSF. A phase screen of zero radians was propagated through the aperture in the same manner as the reconstructed residual phases. The peak intensity value of a particular frame divided by the peak value of the diffraction-limited image is an estimator of SR in the short-exposure regime. In Figure 1 we show the best and the worst frame in the dataset PSF 2, together with the corresponding phase maps.

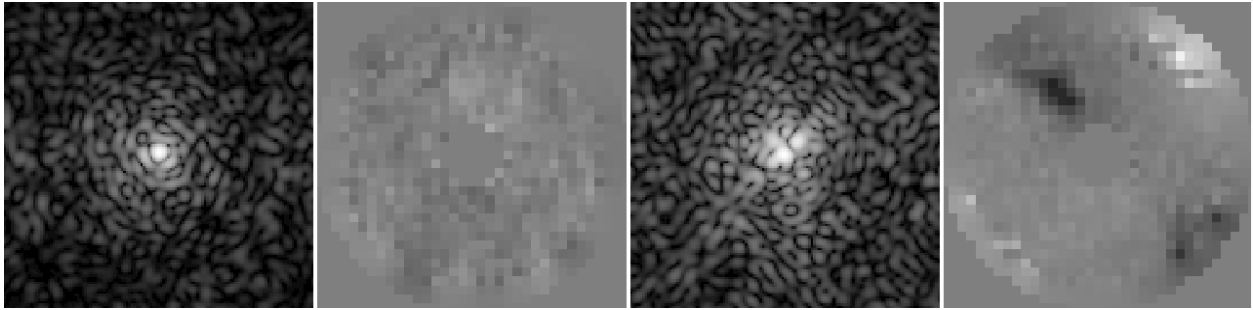

Fig. 1. The best (*left*), and the worst (*right*) frames in the dataset denoted PSF 2, displayed on a logarithmic scale. The Strehl ratios are 0.63 and 0.28 respectively. Also shown are the phase maps, displayed on a linear scale between -10 and 10rad.

The Strehl ratios for the AEOS system reported here are consistently higher than those reported in the past (Roberts & Neyman 2002). This is due to a number of factors. The SR of the science camera (Visible Imager) is 70% and the algorithm used in previous studies underestimated the focal-plane Strehl ratio by 10-15% (Roberts et al. 2004) which reduce the SR at the corrected pupil from 40% to a measured one in the focal plane of 25%. In addition, the system performance has since been improved as has the accuracy of SR measurement and now typical focal plane SR values are in the order of 30-40% corresponding to pupil-plane measurements of 42-57%. We also note that there is a significant residual tip-tilt error so that long-exposure images have a further SR reduction. Thus the Strehl ratio measurements estimated here from numerically propagating the wavefront measured in the pupil plane to a focal plane PSF appear reasonable, when taking into account the SR losses discussed above, for the atmospheric conditions, i.e. $r_0$, of the observations.

The lack of background and readout noise sources meant we could measure SR accurately. On the other hand the numerically-generated PSFs cannot be considered true AO images because of the absence of high-order atmospheric aberrations. This, in turn, translates into the statistics of SR (§ 6). AEOS datasets show more significant negative skewness than the Lick measurements for similar mean SR. The SR dispersion is also artificially reduced. This can be explained by the fact that higher-order aberrations scatter light away from the optical axis and therefore all effects pertaining to the on-axis intensity are slightly overestimated in our AEOS observations.

3. TEMPORAL PROPERTIES OF THE INSTANTANEOUS STREHL RATIO

Figure 2 shows a representative sample of SR and phase variance time series. Trends and peaks existing on very short timescales can be easily identified for both quantities. Sometimes these sudden jumps happen coincidentally for the two time series (PSF 2), on other occasions the correspondence is hard to see (PSF 4) partly because of the large scale of phase variance fluctuations. Correlated changes in the level of correction, especially visible in the phase variance time series for PSFs 2 and 4, support the hypothesis that the turbulence does indeed change on the timescales of 0.05-0.1s (Bradford & Roberts 2007). Because these trends are very rapid AO is not able to compensate them.

Eleven quasi-stationary datasets, as represented by PSF 13, were identified and used for the statistical analysis in sections 5 and 6. The stationary SR datasets had an excess of low values, as can be seen in Figure 2 for PSF 13. The reverse was true for stationary time series of the phase variance.

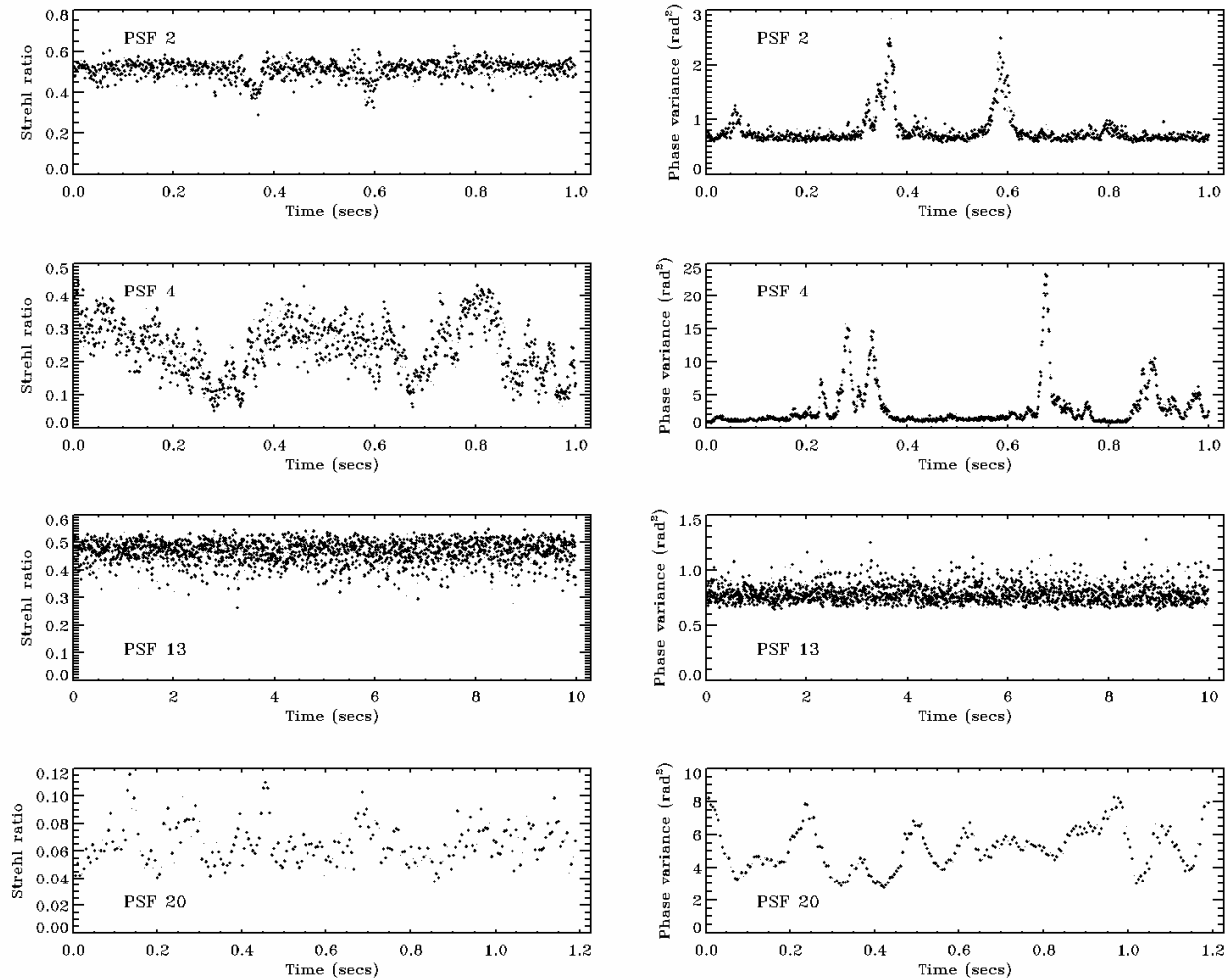

Fig. 2. Temporal sequences of SR and phase variance for four datasets representative of all observations.

4. INVESTIGATION OF THE MARÉCHAL APPROXIMATION

The AEOS data permit investigation of the accuracy of the Maréchal approximation, which relates SR to the phase variance (Born & Wolf 1980). This approximation is usually stated as follows:

$$SR = 1 - \sigma^2 \qquad (1)$$

where $\sigma^2$ denotes the aperture-averaged spatial phase variance and not the ensemble-averaged phase variance. This is an important distinction, because the other form of the Maréchal approximation

$$\langle SR \rangle = e^{-\sigma^2} \qquad (2)$$

was derived for the *statistical* phase variance and the ensemble-averaged SR. Equation (1) was obtained by expanding the electric field in the Taylor series, and it is valid only for very high SR. While the range of validity of equation (2) has been proven to be somewhat larger than for equation (1) the former has been exclusively used for long-exposure SR. There is nothing preventing the use of equation (1) for short exposures with removed tip and tilt. We will now demonstrate that equation (2) can also be used in the case of short exposures, assuming one has access to discrete samples of the phase.

SR is defined here as the ratio of the peak value of the short exposure *I* divided by the peak value of the diffraction-limited image *I\**. This ratio can be rewritten remembering that the focal plane image is the squared modulus of the Fourier transform of the electric field at the entrance pupil. For de-tilted wavefronts the peak value of the image corresponds to the value at the origin of the focal plane, so the Fourier transform reduces to:

$$SR = \frac{I_{peak}}{I^*_{peak}} = \frac{1}{A}\left|\iint_A e^{-i\phi(x,y)} dxdy\right|^2 \tag{3}$$

where x and *y* are variables of integration over the area *A*. The aberrated phase at the aperture is denoted by $\phi(x,y)$.

For the Shack-Hartmann wavefront sensor, one has a discrete set of estimates of the phase, rather than a continuous surface. The integral is then approximated by the sum over the phase estimates:

$$SR = \frac{1}{m^2}\left|\sum_{k=1}^{m} e^{-i\phi_k}\right|^2 = \frac{1}{m}\sum_{k=1}^{m} e^{-i\phi_k} \frac{1}{m}\sum_{l=1}^{m} e^{i\phi_l} \tag{4}$$

where we used the fact that $dxdy = A/m$, *m* is the number of subapertures. Each of the sums in equation (4) is a sample estimator of mean value of $e^{i\phi}$ or $e^{-i\phi}$. Assuming that $\phi$ is normally-distributed (see § 5) equation (4) approximates the following expression

$$SR = \int \frac{1}{\sigma\sqrt{2\pi}} e^{-\frac{(\phi-\bar{\phi})^2}{2\sigma^2}} e^{-i\phi} d\phi \cdot \int \frac{1}{\sigma\sqrt{2\pi}} e^{-\frac{(\phi-\bar{\phi})^2}{2\sigma^2}} e^{i\phi} d\phi \tag{5}$$

where the bar denotes mean value. The integrals in equation (5) represent characteristic functions of the Gaussian random variables, evaluated at the value of one. The solution to the last equation can be found in standard textbooks on the probability theory:

$$SR = e^{-\sigma^2/2} \cdot e^{-\sigma^2/2} \tag{6}$$

which of course reduces to equation (2). The important aspect here is that we have to invoke ergodicity in order to use spatial phase variance. This is a good assumption since part of our data reduction is the removal of static aberrations from the wavefronts. These static aberrations introduce spatially-varying mean value.

Since we are interested in shift-invariant quality metrics, our goal was to check the approximation for de-tilted wavefronts and peak intensities. Figure 3 shows the predicted and observed SR.

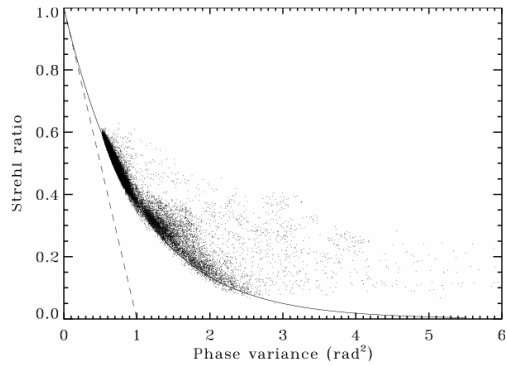

Fig. 3. The Maréchal approximation tested on the de-tilted wavefronts and instantaneous SR: dashed line corresponds to equation (1), solid line to equation (2).

Two things are immediately noticeable in Figure 3. Firstly, equation (2) provides a much better fit to the observed SR than equation (1). Secondly, both forms of the Maréchal approximation underestimate SR. It can also be observed that the accuracy of equation (2) decreases for larger phase variances.

Both forms of the approximation are sometimes used in the same context. This is dictated by the observation that equation (1) contains two terms in the expansion of equation (2) (Tyson 2000). In § 6 we use the Maréchal approximation to explain the shape of the histograms of SR. We decided to employ equation (2) as it gives a significantly more accurate relationship between the phase variance and SR. The average deviation from its prediction was computed to be 2.3%. This shows that equation (2) could be used in the moderate and high-correction regime for the instantaneous SR defined as the peak intensity of a single short exposure.

5. DISTRIBUTION OF THE ESTIMATE OF THE WAVEFRONT PHASE VARIANCE

5.1. Proof for the Gamma Distribution of the Phase Variance

The AEOS data gave us a unique opportunity to estimate the distribution of the phase variance. Figure 4 shows the histograms of the phase variance for four quasi-stationary time series. The shape of these four histograms is representative of the statistics of the eleven stationary sequences. It is clear that the distribution possesses a positive skewness, i.e. an excess of high values is present.

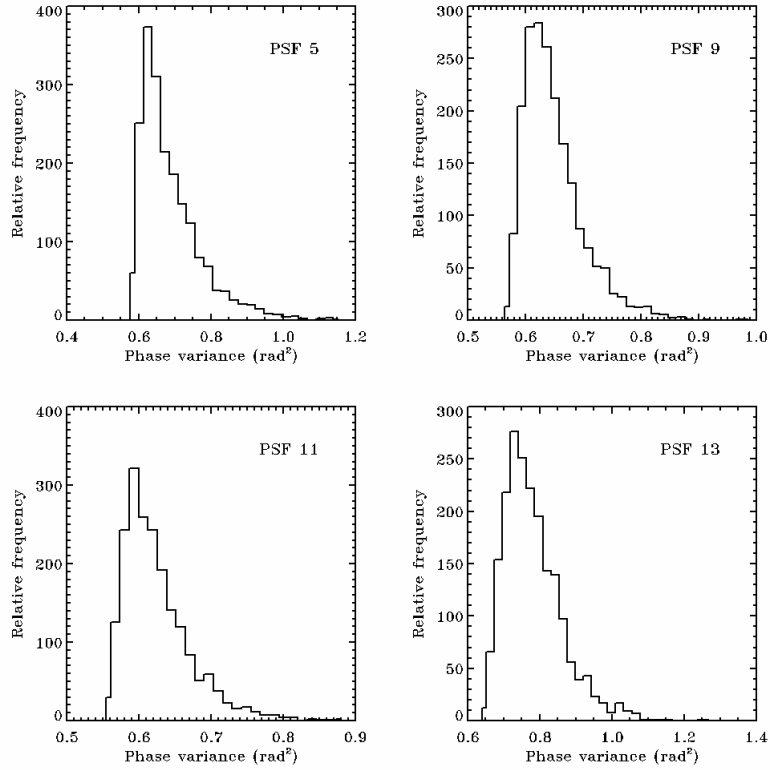

Fig. 4. Histograms of the phase variance for the quasi-stationary time series denoted PSF 5, 9, 11 and 13. Note different scales for the horizontal axes.

The following is a proof that the sample variance calculated from the estimated wavefronts can be treated as gamma-distributed. The notion of "sample variance" relates to the estimate of the *true* variance of the continuous wavefront, which cannot be measured.

Phase perturbed by the atmosphere, $\phi(x,y,t)$, is a Gaussian random process by the central limit theorem (CLT) (Herman & Strugala 1990). The accumulated optical phase difference at the telescope pupil, located at the position $z_0$, is given by

$$\phi(x, y, z_0, t) = \frac{2\pi}{\lambda} \int_z \delta n(x, y, z', t) dz' \qquad (7)$$

where $\lambda$ is the wavelength of the beam, the beam is taken to propagate in the $z$ direction, and optical turbulence is caused by the fluctuations in the refractive index $\delta n$. Since $\phi(x,y,z_0,t)$ is essentially a sum over many $\delta n$'s it will have Gaussian statistics (by CLT).

The AO system can be thought of as a high-pass spatial filter acting on the phase (Sivaramakrishnan et al. 2001). In the simulations performed by Sivaramakrishnan et al. the Fourier transform, denoted by $F$, of the phase is multiplied by the Fourier representation of the high-pass filter, and this product is inverse-transformed to produce the AO-corrected phase:

$$F(\phi_{AO}(x, y, t)) = F(\phi(x, y, t)) \cdot F(f(x, y)) \qquad (8)$$

where $f(x,y)$ is a spatial representation of the filter and the dependence of phase on the vertical dimension $z$ was dropped. By the convolution theorem the above equation can be represented as a convolution in the spatial domain:

$$\phi_{AO}(x, y, t) = \int_{-\infty}^{\infty} \int_{-\infty}^{\infty} \phi(x', y', t) \cdot f(x - x', y - y') dx' dy' \qquad (9)$$

Following Goodman (2000), this integral can be re-written as a limit of approximating sums:

$$\phi_{AO}(x, y, t) = \lim_{\Delta x', \Delta y' \to 0} \sum_{k=-\infty}^{\infty} \phi(x'_k, y'_k, t) \cdot f(x - x'_k, y - y'_k) \Delta x' \Delta y' \qquad (10)$$

where $x'_k$ represents a point in the middle of an interval $\Delta x'$. Now, $f(x - x'_k, y - y'_k)$ is simply a known real number, so the above formula is actually a weighted sum of the normally-distributed atmospheric phase values. The sum of any number of Gaussian random variables, dependent or independent, is itself Gaussian. Hence the AO-corrected phase is a spatio-temporal Gaussian random process.

The aperture-averaged spatial phase variance is given by:

$$\sigma_\phi^2 = \frac{4}{\pi D^2} \int w(x, y) [\phi(x, y, t) - \bar{\phi}(x, y, t)]^2 \, dx dy \qquad (11)$$

where $w(x,y)$ is the telescope pupil function:

$$w(x, y) = \begin{cases} 1, & \text{if } \sqrt{x^2 + y^2} < D/2 \\ 0, & \text{otherwise} \end{cases} \qquad (12)$$

One never has access to the true value of the phase variance; rather it can be estimated at the locations of the WFS subapertures or actuators. It can be assumed that the estimated phase values $\hat{\phi}_{AO}(x_i, y_i, t)$ constitute a Gaussian sample since they approximate values of the Gaussian sample $\phi_{AO}(x_i, y_i, t)$, i.e. AO-corrected phase at the actuator locations $x_i$, $y_i$.

There is also another argument. Values of $\hat{\phi}_{AO}(x_i, y_i, t)$ are obtained via multiplication of the reconstruction matrix – with dimensions equal to the number of actuators and twice the number of subapertures – and the centroids vector. In the case of the AEOS system this means that each estimated phase value is computed as a weighted sum of a great number, $\cong 950$, independent random variables (centroids). The possible dependence of centroids is not a problem as they are Gaussian random variables (see previous discussion of the Gaussian AO-corrected phase). By the CLT, the resulting estimates are normally distributed.

In this paper the aperture-averaged phase variance $\sigma_\phi^2$ is estimated using the maximum likelihood estimator:

$$\hat{\sigma}^2 = \frac{1}{m} \sum_{i=1}^{m} \left( \hat{\phi}_i - \overline{\hat{\phi}} \right)^2 \qquad (13)$$

where $\hat{\phi}_i$, $i = 1,\ldots,m$, are the estimated phase values and $\overline{\hat{\phi}}$ is their mean. This sample variance follows a gamma distribution ($\Gamma$) if all $\hat{\phi}_i$ are independent and identically distributed (i.i.d.) (Kenney & Keeping 1962):

$$p(\hat{\sigma}^2) = \Gamma\left( \frac{m-1}{2}, \frac{2\sigma_\phi^2}{m} \right) \qquad (14)$$

This is a sampling distribution of the phase variance. It arises because of the finite sample size. The true phase variance $\sigma_\phi^2$ could be constant, while the variance estimated from the actuator commands would vary according to the gamma distribution.

The PDF of the three-parameter gamma distribution is given by the formula:

$$p(x;k,\theta,\mu) = \frac{\left(\frac{x-\mu}{\theta}\right)^{k-1} \exp\left(-\frac{x-\mu}{\theta}\right)}{\Gamma(k)\theta} \quad \text{for } x \geq \mu \tag{15}$$

where $k > 0$ is the shape parameter, $\theta > 0$ is the scale parameter, and $\mu$ is the location parameter, which shifts the PDF left and right. Equation (10) suggests $k = (m-1)/2$, and $\theta = 2\sigma_\phi^2/m$. The gamma function, $\Gamma(x)$, is given by:

$$\Gamma(x) = \int_0^\infty t^{x-1} e^{-t}\, dt \tag{16}$$

The PDF is positively skewed, its mean is equal to $\mu + k\theta$ and variance is $k\theta^2$. Both exponential and $\chi^2$ distributions are special cases of the gamma distribution. Since the location parameter simply shifts the PDF it is usually omitted in the definition, although it is useful in the fitting process.

The gamma distribution (for $\mu = 0$) has the property of constant coefficient of variation, defined as the ratio of the standard deviation to the expected value of a random variable. For the two-parameter gamma distribution it is equal to $k^{-1/2}$. This means that during periods of poor seeing/poor compensation phase variance will be more variable as the mean deviation scales linearly with the mean level of fluctuations. This effect will be discussed in the next section.

5.2. Heteroskedasticity of the AEOS Phase Variance

The AEOS observations (both the phase variance and the Strehl ratio) display large heteroskedasticity, i.e. the variance of the data depends on the mean level. It is hypothesized that the phase variance deviations from the mean scale linearly with that mean, as is the case with a gamma-distributed random variable. In order to show this effect, one has to gain access to the residuals. One approach is to use B-splines (Green & Silverman 1994), where an unknown smooth function is approximated as the weighted sum of the basis polynomials. B-splines are often used for the purpose discussed here (e.g. Smith et al. 2003).

To illustrate the change in variance one can plot the residuals vs. the expected values. This is called the scale-location plot. If the coefficient of variation $CV$ is constant then:

$$\sqrt{\sigma_X^2} = CV \cdot \overline{X} \tag{17}$$

where $X$ is the measured random variable – phase variance in this case, and $\sigma_X^2$ is its variance. After taking logarithms of both sides one gets:

$$\log\left(\sqrt{\sigma_X^2}\right) = \log(CV \cdot \overline{X}) = \log(CV) + \log(\overline{X}) \tag{18}$$

This means that on a log-log scale-location plot, a linear relationship with a unit slope should be easy to identify. The spline model at every point is taken to represent the expected values, and the absolute value of the residuals is taken to be a good estimate of the standard deviation. Figure 5 presents log-log scale-location plots for two highly variable phase variance time series (PSFs 2 & 4). Linear fits are also shown. Heteroskedasticity is evident. It was observed that *all* non-stationary series exhibit this behaviour.

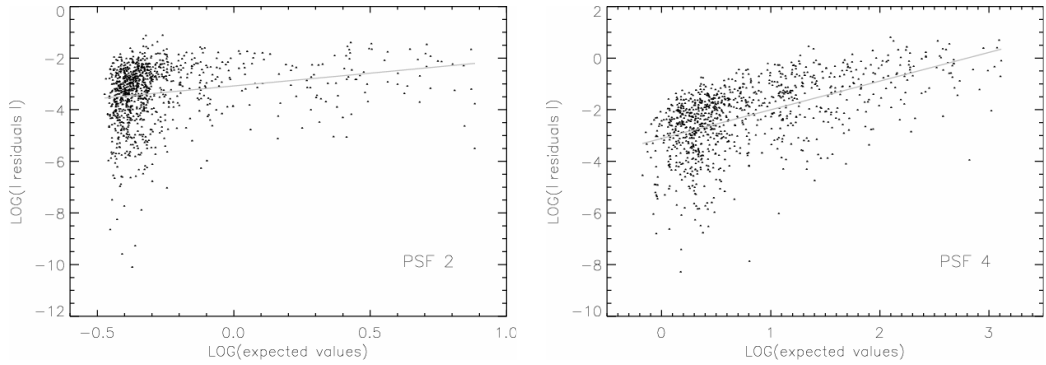

Fig. 5. Scale-location plots for the phase variance in PSFs 2 and 4, on logarithmic axes.

The slopes obtained via linear regression were 0.96 and 1.11 for PSF 2 and 4 respectively. This is very close to the predicted value of one – see equation (18). From the fit coefficients we also extracted the number of independent phase samples $m$. This can be done because $CV^2 = (m-1)/2$. It is interesting to note that the obtained numbers, 945 and 1015, are very close to the number of un-obscured subapertures, which equals 950.

With heteroskedastic samples, the usual practice is to use a variance-stabilizing transformation (Yamamura 1999), most often by taking logarithms of the original time series. This particular transformation is very useful when the hypothesized distribution is log-normal, because the goodness-of-fit tests can be subsequently performed using the Gaussian PDF. Transforming the AEOS data does lead to stable variance, but is detrimental in light of the goodness-of-fit testing to follow.

It should also be said that SR sequences displayed the opposite trend to phase variance, i.e. the dispersion decreased for higher mean SR. This aspect is treated more extensively in § 6.

5.3. Goodness-of-Fit Tests

To test the proposed gamma model for phase variance one has to reduce the data to i.i.d. residuals. It proved impossible for the non-stationary sequences due to the heteroskedasticity. The eleven quasi-stationary series were successfully fitted with the appropriate models and the residuals were extracted. This was done using the ARIMA modelling (Box & Jenkins 1970). ARIMA stands for AutoRegressive Integrated Moving Average. It is a statistical method of discovering patterns in data and forecasting future values based on those patterns. It was used here for the opposite goal – to obtain the values of the i.i.d. "noise" and test its distribution. Mixed ARIMA models of order 2 were often sufficient to ensure non-significant autocorrelation of the residuals. Table 2 lists the stationary sequences with their respective ARIMA models.

As a visual example we show the histogram of the residuals of PSF 13 fitted with the three-parameter gamma PDF given by equation (15). This is shown in Figure 6 which indicates that the gamma PDF provides a very good fit to the distribution of phase variance obtained from the estimated wavefronts.

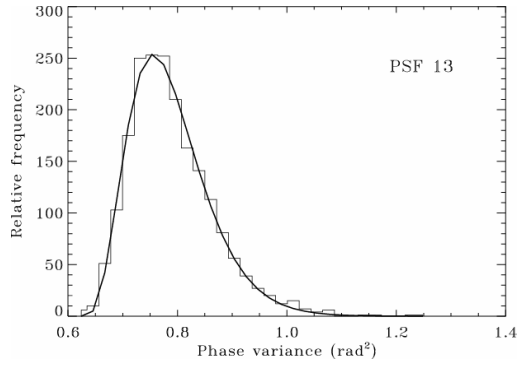

Fig. 6. Fit of the gamma distribution to the histogram of the phase variance for HR 6872 (PSF 13). The residuals of the ARIMA model are centred on the series mean.

This is only a visual test. The Kolmogorov-Smirnov goodness-of-fit test (Wall & Jenkins 2003) was used to obtain quantitative confidence levels for the null hypothesis that a sample comes from the gamma distribution. The test statistic $D$ is the maximum of the absolute difference between the empirical distribution function (EDF) – which is the proportion of the observed values that are less or equal to a particular value – and the hypothesized cumulative density function (CDF). If the null hypothesis – that CDF is the underlying distribution – is correct, EDF should be close to CDF and $D$ should be close to zero. The result of the test, the $p$-value, gives the probability that a value of $D$ at least as large as the one observed, would have occurred if the null hypothesis were true. The greater the $p$-value the more confidence one can have in the null hypothesis.

When all the parameters of a hypothesized CDF are specified a priori there exists an approximate formula for the computation of the $p$-value. Unfortunately this was not the case here and the bootstrap simulation (Ross 2001; Cheng 2001) had to be used instead. The parameters of the gamma distribution were estimated using the method of moments (Kenney & Keeping 1962). In this approach, sample moments are equated to the unobservable population moments. Then the equations relating the distribution parameters to the population moments are solved. In the case of the three-parameter gamma PDF, parameters $k$, $\theta$ and $\mu$ were estimated using the following set of equations (Wilks 2000):

$$k = \left(\frac{2}{m_3}\right)^2$$
$$\theta = \frac{\sigma \cdot m_3}{2} \qquad (19)$$
$$\mu = \overline{X} - \frac{2\sigma}{m_3}$$

where $m_3$ is the skewness of a random variable and $\sigma$ is its standard deviation. The values of these moments were estimated from a sample and substituted in the above equations. The CDF of the gamma distribution is expressed in terms of the incomplete gamma function $\gamma(a,x)$:

$$F_X(x) = \frac{\gamma\left(k, \frac{x-\mu}{\theta}\right)}{\Gamma(k)} \qquad (20)$$

where the gamma function $\Gamma(x)$ is given by equation (16), and the (lower) incomplete gamma function is

$$\gamma(a,x) = \int_0^x t^{a-1} e^{-t}\, dt \qquad (21)$$

The sample parameters $\hat{k}$, $\hat{\theta}$ and $\hat{\mu}$ were plugged into equation (20). The value of the CDF was computed for each phase variance measurement in a dataset. The maximum distance between the EDF and the CDF was then found.

The bootstrap simulation is a Monte-Carlo approach to estimating the confidence intervals for the null hypothesis. The random variables with the hypothesized distribution are generated, and the parameters of the distribution are estimated from the simulated samples using the same method as for the observed sample. Again, the test statistic *D* is calculated as the maximum distance between the EDF and the CDF. The result of the bootstrap method is the number of times the test statistic *D* calculated from the generated sample is greater or equal to *D* calculated from the observed sample. This number divided by the total number of simulations gives the *p*-value.

Gamma-distributed random variables were generated using the acceptance-rejection algorithm (Press et al. 1992). Ten thousand samples of the same size as a given dataset were generated. *p*-values are listed in Table 2. In all cases we obtained values higher than 0.5. This indicates that the three-parameter gamma PDF provides a remarkably good model for the estimated phase variance.

Table 2 Results of the goodness-of-fit tests applied to the stationary phase variance time series. Corresponding ARIMA(p, d, q) models are also listed, where p is the order of autoregression, d is the order of differencing, and q is the order of moving-average involved.

| PSF No. | Bright Star Catalogue ID | ARIMA model (p, d, q) | K-S p-value |
|---|---|---|---|
| 5 | HR 6713 | (2, 0, 2) | 0.6 |
| 6 | HR 7942 | (2, 0, 0) | 0.95 |
| 7 | HR 6713 | (2, 0, 2) | 0.52 |
| 8 | HR 6713 | (1, 0, 2) | 0.8 |
| 9 | HR 6008 | (2, 0, 2) | 0.57 |
| 10 | HR 6008 | (2, 0, 2) | 0.73 |
| 11 | HR 6498 | (2, 0, 1) | 0.71 |
| 12 | HR 6498 | (2, 0, 1) | 0.82 |
| 13 | HR 6872 | (2, 0, 3) | 0.83 |
| 14 | HR 7744 | (2, 0, 1) | 0.7 |
| 15 | HR 7744 | (2, 0, 1) | 0.9 |

6. STREHL RATIO DISTRIBUTION BASED ON THE PHASE VARIANCE PDF

Negatively-skewed distribution of SR was first observed in our Lick campaign (Gladysz et al. 2006). Similarly to the Lick measurements, the AEOS stationary SR histograms have low-end tails (Figure 7). The negative skewness is very significant in these data. We also observed positive skewness in the low-mean time series, but these sequences were non-stationary. Therefore the histograms in the bottom panel in Figure 7 can only be treated as rough approximations to the underlying theoretical PDF.

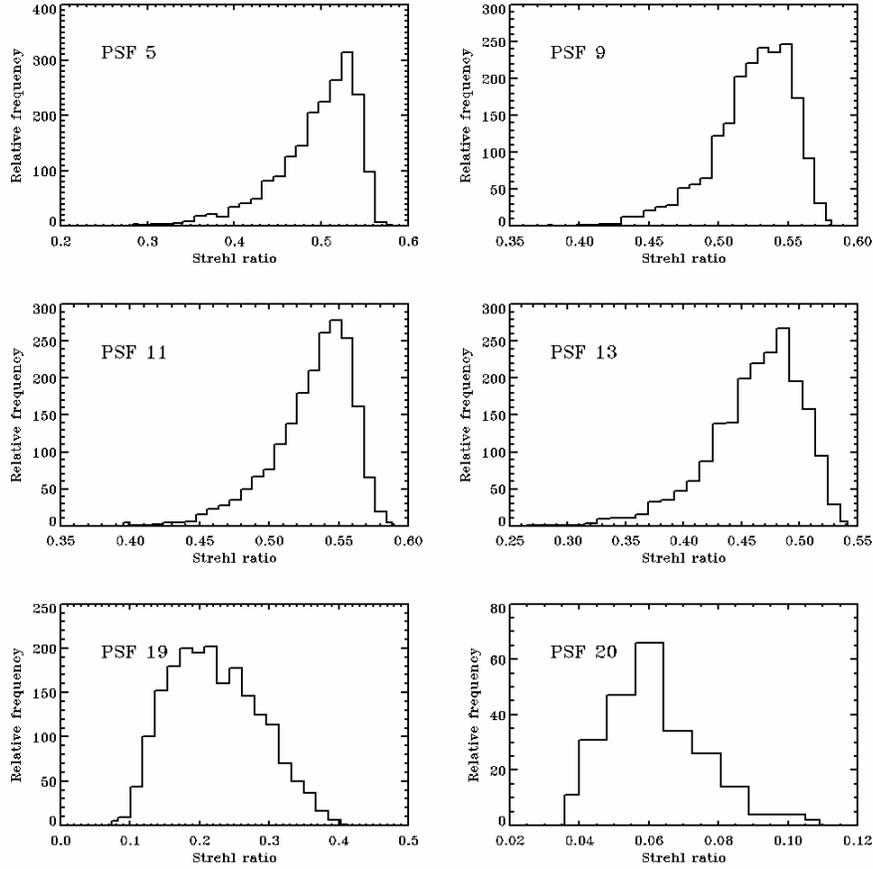

Fig. 7. SR histograms for four quasi-stationary time series, and two non-stationary, low-mean sequences (PSFs 19 and 20). Note different scales of the abscissas.

We explain this effect by transforming the PDF of the phase variance. If there exists a monotonic functional relationship between two random variables, and the distribution of the first variable is known, then this known PDF can be recast to obtain the PDF of the second variable (Goodman 2000). Here, the proposed gamma model for the phase variance estimate, together with equation (2) is used to arrive at the simple result:

$$p_{SR}(sr) = \frac{p_{\hat{\sigma}^2}(-\ln sr)}{sr} \qquad (22)$$

where $SR$ denotes the random variable with possible values $sr$ and $p_{\hat{\sigma}^2}(\ )$ is the distribution of the phase variance estimate described by equations (14) and (15). Recently, Soummer & Ferrari (2007) independently developed similar explanation to the phenomenon of negatively-skewed statistics of high SR.

It is now possible to look at the distribution of the instantaneous Strehl ratio, given the number of phase estimates $m$ (number of subapertures in the WFS) and the true phase variance $\sigma_\phi^2$. Parameters of the system and the atmosphere ($m$ and $\sigma_\phi^2$) are related to the parameters of the gamma distribution $k$ and $\theta$ via equations (14) and (15).

Firstly, the influence of the level of correction ($m$) on the SR distribution was tested. $\sigma_\phi^2$ was decomposed into the fitting error, the bandwidth error, and the error due to pure time delay similarly to the earlier analysis (Gladysz et al. 2006). The increasing number of subapertures (and actuators) only affects the first of these errors. The resulting PDFs are shown in Figure 8.

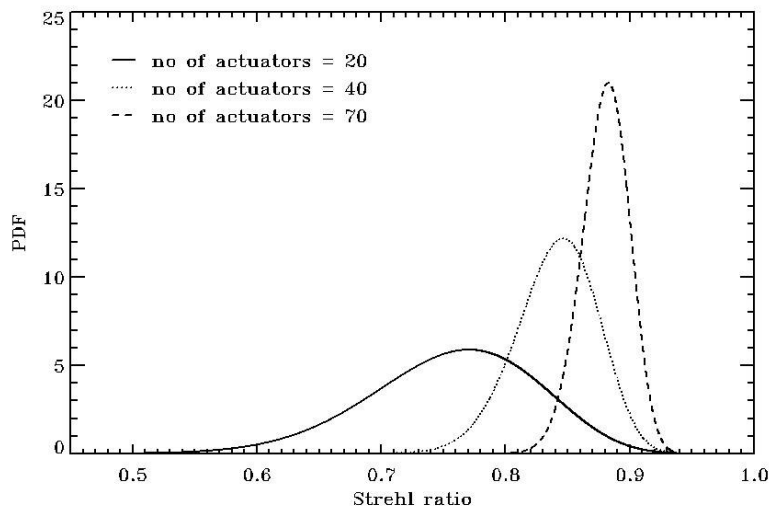

Fig. 8. The distributions of the instantaneous Strehl ratio for three levels of the actuator density.

One should notice that the distributions are negatively skewed as expected. Also the variance decreases for higher *m*. We do not have same-wavelength data from the Lick and AEOS systems to perform verifications of this prediction.

Given equation (22) it is also possible to check what happens to the distribution of SR when the level of correction (*m*) is constant but the mean turbulence strength ($r_0$) is changing. The number of subapertures was set to 35. Figure 9 shows the results.

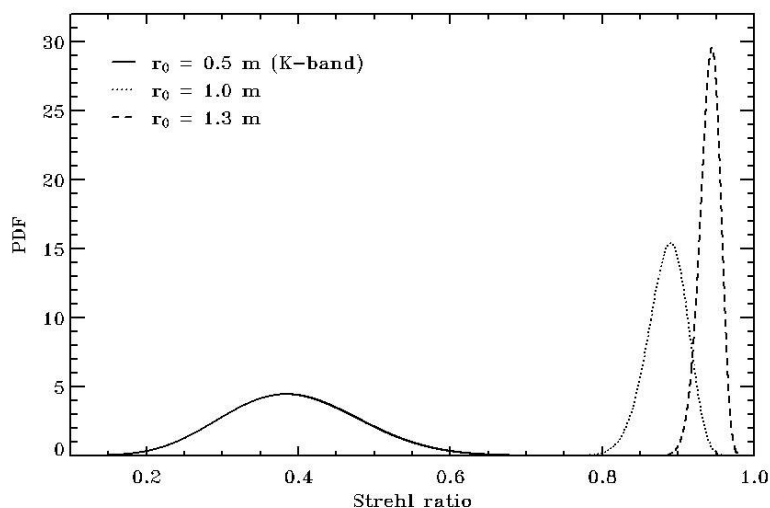

Fig. 9. Distributions of the Strehl ratio for three levels of turbulence strength when the number of subapertures is constant.

Again, the variance of estimates decreases for higher mean SR (longer coherence lengths). Note that the skewness changes from positive to negative around SR = 0.4. Skewness shifts towards more negative values for higher means.

Figures 8 and 9 imply that for higher mean SR two effects should be taking place: decreasing variance and decreasing (negative) skewness. These effects were observed in the AEOS datasets. It should be noted that the differences in mean SR between datasets was more often produced by the differences in brightness of the observed objects, rather than changes in atmospheric turbulence (Table 1 shows that two successive stationary observations of the same star resulted in almost identical SR, so $r_0$ was similar for both sequences). In Figure 10 (*left*) there is a clear decreasing trend in SR variance, with three outliers present. The decreasing trend is not as obvious in the skewness plot. The outlying points in the variance plot are again outliers, in addition the last data point, corresponding to mean SR of 0.57, is not significant because this particular dataset had only 400 values, and the sample skewness estimator is known to produce unreliable values for small samples.

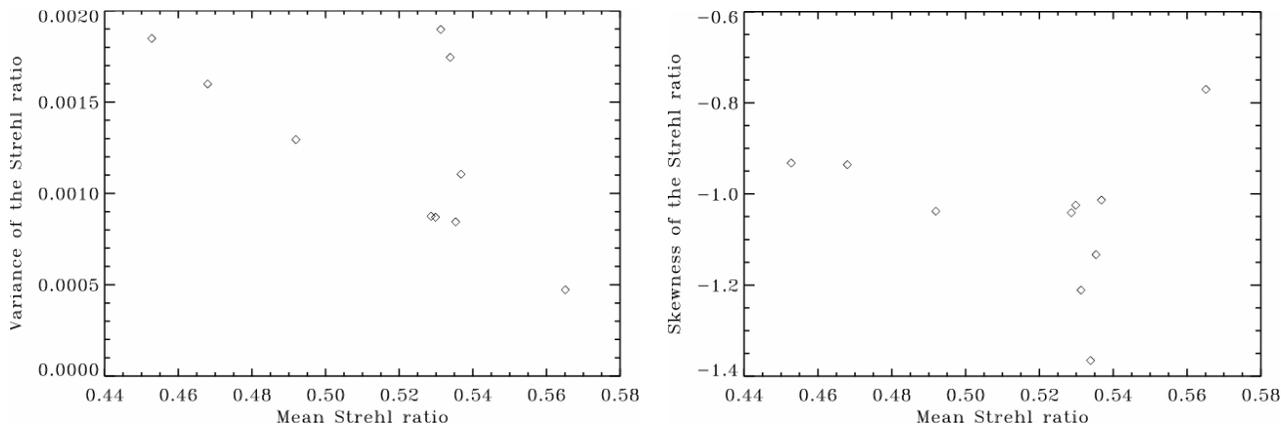

Fig. 10. Variance of the instantaneous Strehl ratio vs. mean value (*left*); skewness vs. mean Strehl ratio (*right*). Only eleven quasi-stationary datasets were used to construct these plots.

We did observe the shift from positive to negative skewness in SR (Figure 7), but the positively-skewed datasets were non-stationary (as is usually the case in the low-compensation regime). This lowers the reliability of PDF estimation via histogram.

The predicted statistics of SR shown in Figures 8 and 9 were tested by scaling of the AEOS estimated wavefronts. Dataset PSF 12 was used. The chosen scaling factors were 0.5, 0.9 and 1.75. SR was computed on the generated images as described in § 2. The histograms of SR were subsequently obtained. Phase variance was also calculated for each scaled wavefront. The parameters $k$, $\theta$, and $\mu$ were computed for the obtained sequence of phase variance using equation (19). Then the predicted PDF for SR was found with the help of equations (15) and (22). The observed and predicted distributions are shown in Figure 11. It can be seen that the predicted PDF approximates the observed one very closely in the moderate to high-SR regime. For low SR equation (2) significantly underestimates the true value. It is interesting to note that a simple scaling of the AO-corrected wavefronts changes the shape of the resulting SR distribution.

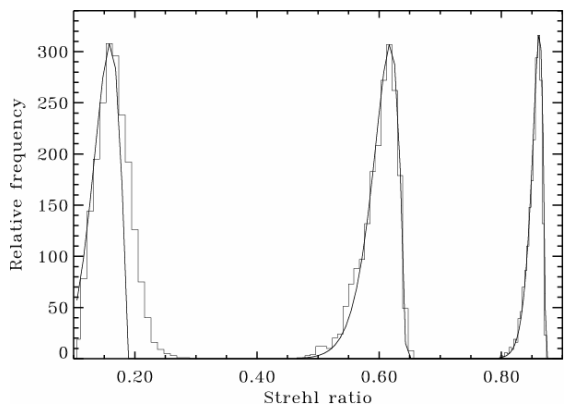

Fig. 11. Predicted and observed distributions of the Strehl ratio when the estimated wavefronts are multiplied by factors: 1.75 (*left*), 0.9 (*centre*), and 0. 5 (*right*).

The above analysis refers to the estimate of SR. But what can be said about *true* SR? We set out to find the distribution of an image quality metric which could be used in the frame selection algorithm. As the name implies, this metric has to accurately describe the quality of a given frame. There are reasons to believe that the PDF of true SR has very similar form to the one shown in Figures 8 and 9.

Firstly, the histograms of SR for the Lick and AEOS data were computed on the images (and not on the wavefronts using equation (2)). Still, the effects outlined above (decreasing variance and skewness shifting towards more negative values for higher means) were observed. Secondly, Monte-Carlo AO simulations with no readout noise and high spatial sampling of the focal plane reproduced the observed PDFs. Also, for the observed time series, SR correlates very well with sharpness defined as the sum of the squared pixel irradiances. This implies that the measurements of SR capture quick changes in the image quality.

There is also another argument utilizing PDF of the true phase variance given by Calef et al. (2005). This PDF has a very complicated equation but the important feature is its significant positive skewness visible in the plot in the paper. The transformation similar to equation (22) is impossible, but in the high-SR regime, where equation (1) is a very good approximation, one can simply take the phase variance PDF and "flip it" to get the negatively-skewed SR PDF.

Given the statistics of our chosen image quality metric we formulate guidelines for the use of frame selection on AO-corrected frames. Our analysis gives reasons to believe that frame selection will produce better relative improvements when seeing is bad or level of correction is modest. In that case the SR has small negative skewness and large variance giving relatively large number of high-quality outliers. Moving to high-correction regime will produce diminishing benefits.

7. CONCLUSIONS

In this paper we presented some methods of establishing the statistics of phase variance and the instantaneous Strehl ratio defined as the peak intensity of a single frame. Models of the statistics of phase variance and SR were put forward and shown to agree well with the data. In particular, the distribution of SR has significance in the field of high-temporal resolution imaging. This distribution can be used to predict the performance of frame selection in different atmospheric conditions and in different science tasks (Gladysz et al. 2007). We briefly summarize these cases here, while encouraging the reader to consult the aforementioned publication for more details.

"Lucky imaging" is traditionally used as a way to enhance the resolution of astronomical images (Law et al. 2006). SR is proportional to resolution and maximizing the former will lead to an improvement in the latter. For the observation of closely-spaced unresolved double stars one can utilize frame selection to try to gain resolution. Seeing that in the low-SR regime the distribution of SR has relatively large variance and positive skewness we postulate that one can always increase resolution by applying frame selection (at the expense of the signal-to-noise ratio). Above SR of approximately 0.5 the distribution becomes negatively-skewed and the method would produce diminishing results.

Frame selection can also be used as a method for the detection of faint companions. This novel application of the old method is helped by the fact that the observer is struggling against quasi-static speckle noise in this scenario. Temporal integration, i.e. accumulation of images with constant noise pattern, does not significantly increase detectability as quantified by the signal-to-noise ratio. Again, knowledge of the SR PDF allows for modelling of the frame selection's results in that case. Surprisingly, one can increase both: SR and the signal-to-noise ratio by the careful application of the method. While the PDF produces more high-quality outliers in the bad-seeing (low-correction) regime, the static speckle contribution which can be circumvented by frame selection is relatively less important in that compensation range. Modelling of the method's outcome when applied to high-contrast imaging is left for future work.

This research was supported by Science Foundation Ireland under Grants 02/PI.2/039C and 07/IN.1/I906, as well as the National Science Foundation Science and Technology Center for Adaptive Optics, which is managed by the University of California at Santa Cruz under cooperative agreement AST 98-76783. Our work was also funded by the Air Force Office of Scientific Research and by Air Force Research Laboratory's Directed Energy Directorate under contracts F29601-00-D-0204 and FA9451-05-C-0257. We would like to thank Granville Tunnicliffe-Wilson and Jerome Sheahan for the introduction to time series analysis and Chris Dainty for support for this research.

The authors wish to acknowledge the SFI/HEA Irish Centre for High-End Computing (ICHEC) for the provision of computational facilities and support.